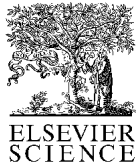
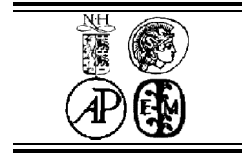

# Quantum, cyclic, and particle-exchange heat engines

T. E. Humphrey,[a,*] H. Linke[b]

[a]*Engineering Physics, University of Wollongong, Wollongong 2522, Australia.*
[b]*Materials Science Institute and Physics Department, University of Oregon, Eugene OR 97403-1274, U.S.A.*

**Abstract**

Differences between the thermodynamic behavior of the three-level amplifier (a quantum heat engine based on a thermally pumped laser) and the classical Carnot cycle are usually attributed to the essentially quantum or discrete nature of the former. Here we provide examples of a number of classical and semiclassical heat engines, such as thermionic, thermoelectric and photovoltaic devices, which all utilize the same thermodynamic mechanism for achieving reversibility as the three-level amplifier, namely isentropic (but non-isothermal) particle transfer between hot and cold reservoirs. This mechanism is distinct from the isothermal heat transfer required to achieve reversibility in cyclic engines such as the Carnot, Otto or Brayton cycles. We point out that some of the qualitative differences previously uncovered between the three-level amplifier and the Carnot cycle may be attributed to the fact that they are not the same 'type' of heat engine, rather than to the quantum nature of the three-level amplifier per se.

## 1. Introduction

'Quantum' heat engines utilize effects such as discrete energy levels [1-8], quantum coherence [5-8] or quantum confinement [9-11] in the process of obtaining useful work from a temperature differential. A number of very interesting results have recently been obtained, including the extraction of work from a single quantum heat bath [5,6,12], efficiency loss due to quantum friction (dephasing) in the three-level amplifier [3] and due to quantum measurement in the two-level (spin) quantum heat engine [7].

Here we point out a difference in the thermodynamics underlying the three-level amplifier and two-level quantum heat engines, which is important for correctly interpreting the qualitative thermodynamic behavior of the three-level amplifier. The two-level quantum heat engine consists of a working gas of non-interacting two-level systems that undergoes a cyclic process involving alternate interaction between a hot and cold reservoir. Reversibility is approached when the *well-defined* temperature of the working gas is the same as the reservoir it is in thermal contact with, i.e. when the heat transfer is isothermal [8]. The two-level quantum heat engine may be therefore be interpreted as a quantum analogue of classical cyclic heat engines.

---

[*] Corresponding author. Tel.: +61 2 9874 0928; Web: www.humphrey.id.au; e-mail: tammy.humphrey@unsw.edu.au.

The three-level amplifier [1-4], on the other hand, is a quantum heat engine that should not be interpreted as a quantum analogue of classical cyclic heat engines, as has been occasionally suggested [2,13]. It is the central thesis of this paper the three-level amplifier is a quantum mechanical analogue of other semiclassical and classical heat engines such as thermionic, thermoelectric and thermophotovoltaic devices. In support of this thesis, we first highlight the fundamentally different thermodynamic mechanisms utilized by classical 'cyclic' and what we will here term 'particle-exchange' (PE) heat engines to achieve reversibility and finite power. In the next several sections we show that the three-level amplifier, photovoltaic [14,15], thermionic [16] and thermoelectric [17] devices, as well as a classical 'toy' gravitational heat engine, are all PE heat engines that approach reversibility in the same way, via isentropic but not isothermal particle transfer between hot and cold heat reservoirs.

## 2. Cyclic heat engine

Cyclic heat engines utilize a working gas that moves through a reversible cycle to transfer heat between hot and cold heat reservoirs and do useful work [18]. A working gas may be defined as a system that is at all times close to thermal equilibrium, so that it has well-defined state variables such as temperature [18]. A diagram of a generic cyclic heat engine is shown in Fig. 1. The processes A-B, B-C, C-D and D-A (in principle there may be a different number of steps) depend upon the details of the specific embodiment of the cyclic heat engine. Examples of classical cyclic heat engines include the Carnot cycle, which consists of two adiabatic and two isothermal steps, the Otto cycle, which consists of two isochoric and two adiabatic steps and the Joule/Brayton cycle which consists of two isentropic and two isobaric steps [19].

As mentioned in the introduction, the two-level quantum heat engine [7,8] is also an example of a cyclic heat engine, in which the working gas is an ensemble of non-interacting two-level systems such as spin ½ systems. In this case steps A-B and C-D constitute adiabatic changes in polarization, while B-C and D-C constitute isothermal changes in polarization. Quantum two-level systems are thus a specific, quantum mechanical embodiment of a Carnot cycle in which the extensive parameter varied during the steps is the average polarization. Note that the population of spins in a quantum two-level heat engine has at all times a well-defined temperature and polarization [8] and so fulfills the requirements for being a working gas.

Maximum efficiency is obtained in cyclic heat engines when the heat exchange between the working gas and the heat reservoir it is in contact with is isothermal [18]. As heat transfer between bodies at the same temperature takes infinitely long, the cycle progresses infinitely slowly and power is not produced at a finite rate. To obtain finite power, it is necessary that the cycle be executed at a finite rate, with non-isothermal heat transfer between the working gas and the reservoirs [19]. One particular situation which has proved relatively easy to analyze and useful for modeling real cyclic heat engines is the endoreversible case [19], in which there is a finite temperature difference between the working gas and the heat reservoir it is in contact with, enabling the transfer of heat in a finite time, while all other aspects of the working gas cycle are assumed to occur without entropy production. The efficiency at maximum power of an endoreversible classical Carnot cycle in which heat transport between the working gas and heat reservoirs is Newtonian, is given by the well-known Curzon-Ahlborn efficiency [20], $\eta_{CA} = 1 - \sqrt{T_C/T_H}$ .

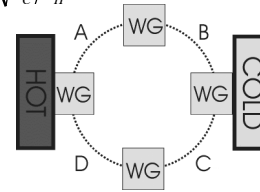

FIG. 1 Schematic of a cyclic heat engine. The essential components are two heat reservoirs and a working gas (WG) that is cycled through a series of quasi-equilibrium states with well-defined temperature.

|  | Cyclic heat engines | PE heat engines |
|---|---|---|
| Essential Components | 1) At least two heat reservoirs at different temperatures<br>2) Working gas<br>3) External work reservoir | 1) At least two particle reservoirs at different temperatures<br>2) Energy selective particle transfer<br>3) Field against which work is done |
| Operation | Cyclic | Continuous |
| Maximum efficiency | Isothermal heat transfer between working gas and heat reservoirs. | Direct, isentropic heat transfer between heat reservoirs via continuous mono-energetic particle exchange. |
| Finite Power | Finite temperature difference between working gas and heat reservoirs. | Transfer of particles with a finite range of energies between heat reservoirs. |

Table 1. Summary of differences between cyclic and particle exchange heat engine.

## 3. Particle-exchange heat engines

Particle-exchange (PE) heat engines may be defined as those in which heat transfer between the hot and cold reservoirs occurs via the exchange of particles in a finite energy range. Work is done against a field by each particle transferred from hot to cold and absorbed from this field for each particle transferred from the cold to the hot reservoir. A diagram of a generic particle-exchange heat engine is shown in Fig. 2.

From a thermodynamic point of view, the most important difference between cyclic and particle exchange heat engines is how they achieve reversibility and finite power. In cyclic heat engines, reversible heat transfer occurs isothermally between the working gas and the reservoirs. In particle exchange heat engines direct heat transfer between the reservoirs occurs isentropically (but non-isothermally) when particles are only exchanged at the energy at which the occupation of states for the particles is the same in both reservoirs.

PE heat engines are characterised by the fact that they would operate with zero efficiency without some restriction on the energy range of particles transmitted between the reservoirs. As a result of this finite energy spectrum, it is important to note that the particles exchanged between the hot and cold reservoirs in PE heat engines do not constitute a working gas, as that state variables such as temperature are undefined. The lack of well-defined state variables for these particles means that the operation of continuous heat engines cannot be represented on a T-S diagram as can be done for the working gas in cyclic heat engines (although this has been attempted for electron heat engines [21]).

Finite power is obtained in PE heat engines when the spectrum of particles transmitted between the reservoirs is finite, and centred above the energy at which the occupation of states is equal for power generation, or below this energy for refrigeration [9-11]. A summary of the differences between cyclic and particle exchange heat engines is given in Table 1.

We now consider in turn thermionic, thermoelectric and photovoltaic devices, and the three-level amplifier, showing that each approaches Carnot efficiency in the same way; via isentropic but non-isothermal monoenergetic particle transfer.

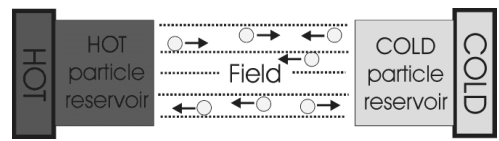

FIG. 2 Principle of a particle exchange heat engine. The essential components are two particle reservoirs at different temperatures, which exchange particles in a finite range of energies and a field against which work is done.

## 4. Ballistic (thermionic) electron heat engines

In an electron heat engine, electrons with a finite range of energies flow between a hot and a cold reservoir against an applied electric field to produce power. In a thermionic device the electron transport between the reservoirs is ballistic (electrons are transmitted between reservoirs without elastic or inelastic scattering events during the transition). In a conventional vacuum thermionic device, the range of electron energies is broad (all electrons with sufficient energy to overcome the potential barrier due to the vacuum are transmitted) [16]. In nanostructured solid-state devices however, where quantum confinement effects alter the electronic density of states and the transmission function, this range can in principle be narrow [9]. Carnot efficiency is achieved in a nanostructured thermionic device when electrons are transmitted only at the energy at which the occupation of states in the two electron reservoirs is the same (marked $E_R$ in Fig. 3) [9]. To show this, we note that the efficiency is given by the work done per electron transmitted from the hot to the cold reservoir, $eV_0 = (\varepsilon_C - \varepsilon_H)$, where $\varepsilon_C$ and $\varepsilon_H$ are the electrochemical potentials in the cold and hot reservoirs respectively, divided by the heat removed from the hot reservoir by each of these electrons, $(E_R - \varepsilon_H)$, so that $\eta_{BE} = (\varepsilon_C - \varepsilon_H)/(E_R - \varepsilon_H)$. By substituting the condition for equal occupation of states in the hot and cold reservoir at $E_R$, where

$$f_H = \left[\exp\left[\frac{(E_R - \varepsilon_H)}{kT_H}\right] + 1\right]^{-1}$$

and

$$f_C = \left[\exp\left[\frac{(E_R - \varepsilon_C)}{kT_C}\right] + 1\right]^{-1},$$

which gives $eV_0 = (E_R - \varepsilon_H)(1 - T_C/T_H)$, it can be shown that Carnot efficiency is achieved.

To obtain finite power from a ballistic electron heat engine it is necessary both to increase the range of energies over which electrons are exchanged, as well as to move the center of this range to either higher (for power generation) or lower (for refrigeration) energies until a net flow of particles from one reservoir to the other is achieved [10].

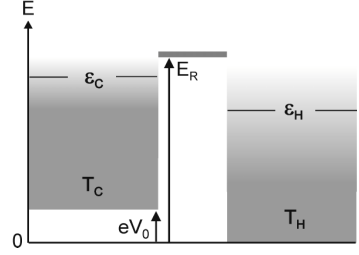

FIG. 3. Schematic of a nanostructured thermionic device. An energy filter such as a resonance in a quantum dot passes electrons between two electron reservoirs at different temperatures and electrochemical potentials at the energy at which the occupation of states in the two reservoirs is the same.

## 5. Diffusive (thermoelectric) electron heat engines

In thermoelectric devices electrons flow diffusively through material that varies continuously in temperature. In another paper [11], we have shown that reversible diffusive electron transport can be achieved if the occupation of electron states is constant across the material at the sole energy, $E_0$, at which electrons are free to move throughout the material (for example if states are limited to a narrow miniband in a quantum dot superlattice, or superlattice nanowire as indicated in Fig 4). This requires that the argument of the Fermi occupation function, $[E_0 - \mu_0(x)]/kT(x)$, is constant as a function of $x$ [11]. In a similar fashion to the ballistic heat engine, we note that the efficiency is given by the voltage across the n-type leg of the idealized thermoelectric nanomaterial shown in Fig. 4 is $\mu_0(L) - \mu_0(0)$ divided by the heat removed from the hot extreme of the n-type leg per electron at open circuit $E_0 - \mu_0(0)$, giving $\eta_{DE} = [\mu_0(L) - \mu_0(0)] / [E_0 - \mu_0(0)]$. If the Fermi occupation function at $E_0$ is equal throughout the material, then $[\mu_0(L) - \mu_0(0)] = [E_0 - \mu_0(0)](1 - T_C/T_H)$, and Carnot efficiency is achieved in the limit that the lattice thermal conductivity of the material tends to zero.

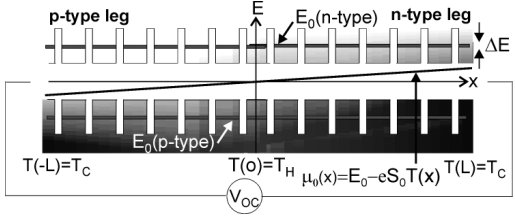

FIG. 4. Reversible thermoelectric nanomaterial such as a quantum dot superlattice or superlattice nanowire, in which the electronic density of states is sharply peaked at one energy $E_0$, and the electrochemical potential for electrons $\mu_0(x)$ varies inversely with the temperature $T(x)$ with a proportionality factor given by the Seebeck coefficient $S_0$ to give a constant occupation of states at $E_0$.

## 6. Solar cells and LEDs

Solar cells and LEDs can be viewed as another embodiment of a particle exchange heat engine. An idealized model is shown in Fig. 5, consisting of a black body radiator at a high temperature (for instance the sun or a hot filament in a heat lamp) and a radiator at a lower temperature (the pn-junction which is the solar cell/LED), which, in this thermodynamic model, exchange photons through an energy filter which can restrict the energy spectrum of exchanged photons to a particular, finite, range (e.g. all energies above the bandgap are transmitted in an ordinary single bandgap solar cell). Photons from the hot body excite electron-hole pairs across the bandgap of the p-n-junction and so generate an electric current that flows against a voltage applied across the junction to produce electrical power [14,15]. An LED operates in reverse, via the application of a voltage that is larger than the open-circuit voltage of the device when operated as a solar cell. In this case, electron-hole pairs arrive at the p-n junction and recombine, emitting a photon which can in principle then be absorbed by the hot black-body and refrigerating pn-junction [22]. In real devices, there are a number of important loss mechanisms such as non-radiative recombination of carriers and the fact that the photons emitted by the p-n junction of a solar cell cannot be considered to be absorbed by the sun [14-15]. However, as is well-known in the photovoltaic community [14,15,22-26], the efficiency of the idealized model shown in Fig. 5 can be shown to be equal to the Carnot limit in the case that the filter only transmits photons with energy equal to the bandgap of the p-n junction.

To show this, we note that the efficiency is given by the work done per photon absorbed form the hot black-body, $eV$, divided by the heat removed from the hot black-body by a photon with energy equal to the bandgap, $E_g$, which gives $\eta_{SC} = eV / E_g$. The occupation of states for photons with the bandgap energy in each of the reservoirs is

$$f_H = \left[\exp\left[\frac{E_g}{kT_H}\right] - 1\right]^{-1}$$

and

$$f_C = \left[\exp\left[\frac{E_g - eV}{kT_C}\right] - 1\right]^{-1},$$

where $eV$ is the chemical potential of photons in the p-n junction [27]. Setting $f_C = f_H$, we obtain $eV = E_g(1 - T_C/T_H)$ and thereby Carnot efficiency. Finite power is obtained when the range of particles exchanged between the blackbodies is increased, with maximum power for the (idealized) solar cell obtained when all photons from the sun with energies greater than the bandgap are transmitted between the hot body and the solar cell.

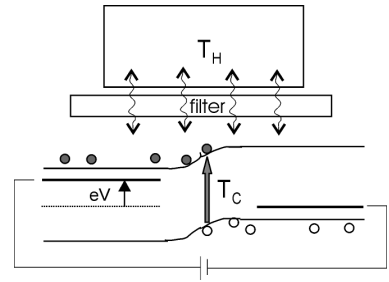

FIG. 5 Model of a solar cell or LED. A hot and a cold black body (the sun and the p-n junction) exchange photons with an energy equal to the bandgap of the p-n junction. When operated as a solar cell, electron-hole pairs generated by the absorption of photons from the hot black-body flow against an applied voltage $V$ to produce electrical power.

## 7. Gravitational ESPE heat engine

Before discussing the three-level amplifier, here we propose a new 'toy' particle exchange heat engine, for the purposes of showing that in principle at least, Carnot efficient PE heat engines are not limited to systems in which particles are quantum or semi-classical, i.e. in which they obey Fermi-Dirac or Bose-Einstein statistics.

The gravitational heat engine, shown schematically in Fig. 6 consists of two or more very thin and wide reservoirs of classical particles, arranged so that the coldest is at a higher gravitational potential energy than the hottest. We assume the existence of energy filters that allow particles in a very narrow energy range to flow freely in either direction between the reservoirs. These could be thought of as a kind of 'even-handed' Maxwell demon that, unlike the usual variety [28], do not attempt to violate the second law. Almost suitable for use as such an energy filter is the 'velocity selector' used by Miller and Kusch [29] to experimentally measure the Boltzmann distribution, which consists of a spinning cylinder with a curved groove cut into it along its length such that particles entering the groove with the correct velocity are transmitted freely. However, particles with the incorrect velocity collide with the walls, and in the original experiment were removed via a vacuum. Therefore, such a device could not be used to realize a *reversible* gravitational heat engine as it either adds energy to the reservoirs via collisions with particles that are returned to the reservoirs, or is a source of energy loss if particles with incorrect energies are removed from the system. It is interesting to note it appears that the only *practical* way to produce a reversible energy filter that does not alter the energy of non-transmitted particles is via resonant tunnelling, an inherently quantum mechanical mechanism.

If a particle with energy $E$ moves from the hot to the cold reservoir, work equal to $mgh$ where $m$ is the mass of the particle and $h$ the height through which is has moved, can be extracted. As with the previous PE heat engines, Carnot efficiency is obtained when the occupation of states in neighboring reservoirs is the same at the one specific energy $E$ at which the hypothetical "energy filter" transmits, i.e. when

$$f_C = \exp\left[-\frac{E - mgh}{kT_C}\right]$$

and

$$f_H = \exp\left[-\frac{E}{kT_H}\right]$$

are equal, so that $mgh = E(1 - T_C / T_H)$. The efficiency is given by $\eta = mgh / E$, and simple substitution of the previous result yields Carnot efficiency.

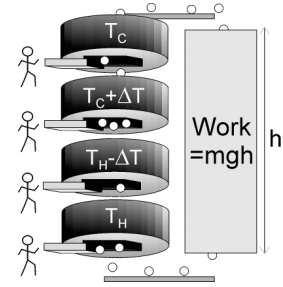

FIG. 6. Conceptual picture of a gravitational heat engine. The temperature differential between the lowest and highest reservoir of particles is used to pump classical particles with a single energy $E$ against the gravitational field. Here the energy filters between the reservoirs are shown as a trapdoor operated by an 'even-handed' type of Maxwell demon that allows particles with energy $E$ (but no others) to pass freely in either direction between the reservoirs (so does not attempt to violate the second law).

## 8. The Three-level amplifier

So far we have provided a number of examples in support our thesis that there exists a class of heat engines that may usefully be distinguished from cyclic heat engines due to the fundamentally different thermodynamic mechanism used to achieve reversibility. As discussed in the introduction, cyclic heat engines utilize a working gas in addition to a hot and cold heat reservoir to achieve isothermal heat transfer and so reversibility, whereas 'particle exchange' heat engines utilize isentropic but *non-isothermal* direct heat transfer via mono-energetic particle exchange between a hot and cold particle reservoir, as summarized in Table 1. We will now discuss the three-level amplifier, and show that it also belongs to this second group of heat engines. We will then conclude by examining the implications of this fact for the study of quantum thermodynamic effects in the three-level amplifier.

The three-level amplifier uses a temperature difference between two blackbody radiators to produce laser light via the exchange of photons, mediated by a three level system [1-4]. A filter between the three-level system and the hot reservoir limits the energy of the photons exchanged between them to $(E_u - E_g)$, as shown in Fig. 7. An electron excited from level g to level u in the three-level system by a photon absorbed from the hot reservoir may then decay in two steps. First from level u to level d, releasing a photon of energy $(E_u - E_d)$ which does work by amplifying a coherent monochromatic radiation field, then from level d to level g, releasing a photon which can be absorbed by the cold reservoir through a second filter which passes only photons with energy $(E_d - E_g)$ [1].

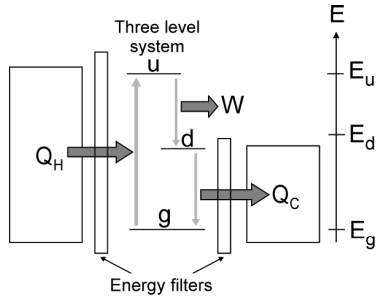

FIG. 7 Schematic of the three-level amplifier. A population of three-level systems absorb photons from a hot black-body via the excitation of an electron from level g to level u. This electron then relaxes by emission of two photons, one at the frequency of a monochromatic coherent radiation field, doing work W, and one that is absorbed by a cold black body through an energy filter at this energy.

As long as the driving field interacts only weakly with the three-level system [3], the three-level amplifier can operate reversibly, with Carnot efficiency, when energy levels of the three-level system are arranged such that the occupation of states at the energy $(E_u - E_g)$ in the hot black body is the same as the occupation of states at the energy $(E_d - E_g)$ in the cold black body [1]. This situation corresponds to the laser being on the verge of population inversion between energy levels u and d. To show that Carnot efficiency is achieved in this case, we note that the efficiency of the laser is the work done against the field per photon removed from the hot bath, divided by the heat removed from the hot bath per emitted photon, $\eta_{TLA} = (E_u - E_d)/(E_u - E_g)$. Population inversion is achieved when the occupation of states at the energy $(E_u-E_g)$ in the hot reservoir,

$$f_H = \left[\exp\left[\frac{(E_u - E_g)}{kT_H}\right] - 1\right]^{-1}$$

equals the occupation of states at the energy $(E_d-E_g)$ in the cold reservoir,

$$f_C = \left[\exp\left[\frac{(E_d - E_g)}{kT_C}\right] - 1\right]^{-1},$$

which yields $(E_u - E_d) = (E_u - E_g)(1 - T_C / T_H)$, giving Carnot efficiency when substituted into the above expression for the efficiency of the three-level amplifier. It can be seen that the three-level amplifier achieves reversibility in the same manner as all of the other PE heat engines considered so far in this paper.

From the perspective of determining what 'type' of heat engine the three-level amplifier is, it is important to note that the three-level system does not constitute a working gas because it is not internally equilibrated; the population of the levels cannot be described by a distribution with a single temperature and chemical potential. In the most in-depth analysis of the 'semiclassical' three-level amplifier (where there is only weak interaction between the three-level system and the field), Geusic et al. [2] dealt with this difficulty by describing the three-level system as a working gas analogous to that in a Carnot cycle and assigning an 'effective' temperature to each of the transitions. As the 'effective' temperature of each transition is the same as that of the reservoir it exchanges photons with if the system is operating reversibly, Geusic et al. propose that the three-level amplifier is a quantum (that is, discrete) analogue of the Carnot cycle as "the cyclic operation consists of isothermal and isentropic interactions". A difficulty with this approach is that it requires an "isothermal" heat exchange between a heat reservoir and a non-equilibrated system that has no well-defined temperature. As noted by Ramsey in his well-known paper [30] "The elements of the thermodynamical system must be in thermodynamical equilibrium amongst themselves in order that the system can be described by a temperature at all".

Here we propose instead to class the three-level amplifier as a 'particle exchange' rather than 'cyclic' heat engine, thus avoiding the above difficulty. This classification has some important implications for interpreting results obtained when quantum effects such as strong interaction between the three-level system and the field, and fields of intense amplitude are introduced, as in the work of Geva and Kosloff [3,4]. As an example, we note that in the high-field limit, Geva and Kosloff have shown that levels u and d each split [4], so that the three-level amplifier effectively consists of two systems that operate in tandem, as indicated in Fig. 8.

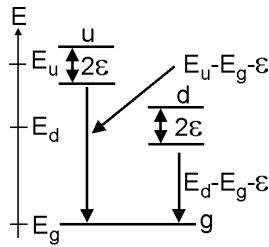

Fig. 8 Illustration of level splitting in the three-level amplifier by an amount equal to the Rabi frequency, $\varepsilon$, in the high-field limit.

As the field amplitude increases the lower heat engine, operating between $u_{low}$, $d_{low}$ and g, switches to refrigeration mode and the two engines work against each other. Eventually the power generation of the upper is exactly balanced by the power consumption of the lower with a net effect that heat is pumped from the hot reservoir to the cold without any generation of work. Geva and Kosloff note that "We are not aware of any similar loss mechanism in classical macroscopic engines" [4], and in another paper [31] Geva states that the three-level amplifier "lacks a well-defined classical analogue". Here we point out that similar behavior, although not of quantum origin, exists in all of the semi-classical and classical PE heat engines discussed in this paper in the situation where the energy filter transmits particles in a finite range around $E_0$, the energy where the occupation of states is equal. In this case transmitted particles with energies greater than $E_0$ generate power while those below $E_0$ consume power, with the result that heat is pumped from the hot to the cold reservoir without the generation of useful work [32].

It would seem therefore that thermionic, thermoelectric and photovoltaic devices, among other more theoretical PE heat engines, provide semiclassical and classical analogues against which quantum PE heat engines such as the three-level amplifier can usefully be compared [33].